\magnification \magstep 1
\hoffset -4truemm
\font\teneul=eufm10
\font\seveneul=eufm7
\font\fiveeul=eufm6
\newfam\eulfam
\textfont\eulfam=\teneul  \scriptfont\eulfam=\seveneul
  \scriptscriptfont\eulfam=\fiveeul
\def\eu{\fam\eulfam\teneul}
\font\tensanserif=cmss10

\def\ssf{\tensanserif}
 2
\parskip=7pt plus 1pt

\def\F{{\cal F}}

\def\e{{\rm e}}

\def\v{{\bf v}}
\def\m{{\bf m}}
\def\n{{\bf n}}
\def\o{{\bf 0}}
\def\p{{\bf p}}
\def\q{{\bf q}}

\def\QED {\hfill\break 
     \line{\hfill{\vrule height 1.2ex width 1.2ex }\quad} 
      \vskip 0pt plus36pt} 
\parskip=14pt
\def\A{{\eu A}}
\def\B{{\eu B}}

\def\Cx{{\bf C}}

\def\Z{{\cal Z}}
\def\Nl{{\bf N}}

\def\Ts{{\bf T}}

\def\Ir{{\bf Z}}

\def\Tr{\mathop{\rm Tr}\nolimits}

\def\idty{{\leavevmode{\rm 1\ifmmode\mkern -5.4mu\else
                                            \kern -.3em\fi I}}}

\overfullrule=0pt

\def\avg{\mathop{\rm Avg}\nolimits}
\def\bavg{\mathop{\hbox{\bf Avg}}\nolimits}
\def\mod{\,\mathop{\rm mod}\nolimits\,}
\def\nl{\hfill\break}
\parindent=20pt

{\nopagenumbers
\font\BF=cmbx10 scaled \magstep 3
\hrule height 0pt
\vskip 3\baselineskip

\centerline{\BF Multi-time correlations} 
\vskip 1\baselineskip
\centerline{\BF in quantized toral automorphisms}
\vskip 2\baselineskip
\centerline{Johan~Andries\footnote{$^1$}{Email: {\tt
             johan.andries@fys.kuleuven.ac.be}}, 
            Fabio~Benatti\footnote{$^2$}{Permanent address: Dept. Theor.
	    Phys. University of Trieste, Italy,} 
            \footnote{}{Email: {\tt
	    benatti@ts.infn.it}}$\!\!$, 
            Mieke~De~Cock\footnote{$^3$}{Onderzoeker FWO, Email: {\tt
	    mieke.decock@fys.kuleuven.ac.be}},
	    Mark~Fannes\footnote{$^4$}{Onderzoeksleider FWO, Email:  
            {\tt mark.fannes@fys.kuleuven.ac.be}}  
            } 
\vskip \baselineskip

\centerline{Instituut voor Theoretische Fysica} 
\centerline{Katholieke Universiteit Leuven} 
\centerline{Celestijnenlaan 200D} 
\centerline{B-3001 Heverlee, Belgium} 
\vskip 3\baselineskip plus 30pt

\centerline{\bf Abstract}
The long time asymptotics of multi-time correlation functions 
of relaxing quantum mechanical systems can be conveniently studied by means of 
free-products of suitable C*-algebras and of states on these free products
given by multiple temporal averages.
In this paper, we study the distribution law of fluctuations
of temporal averages in a class of quantized toral automorphisms.

\vfill\eject}

\pageno=1

\beginsection{1 Introduction}

Recently, it has been suggested~[BF, ABDF] that the statistics of long time
asymptotics of multi-time correlation functions of quantum dynamical systems
might be used to associate particular probability distribution laws with the 
variety of phenomena commonly going under the name of {\ssf quantum chaos}~[CC]. 
The proposed setting is that of a free-product of copies of the algebra of
obervables of the system, equipped with a state obtained by a suitable multiple
time-average.

Given an invariant state and expectations of observables at different
times, with a same time possibly appearing more than once, one may try
to express such expectations in terms of time-ordered correlation
functions. The intrinsic complexity of the dynamics might render this
task very hard indeed and is likely to generate a more or less unwieldy
proliferation of commutators. We will in particular address the
situation where differences between different times become large. In
the particular case of strong time-asymptotic commutativity one can
easily group observables at equal, largely separated, times.

In general, the idea is to leave aside any attempt at simplifying
multi-time correlation functions by expressing them in terms of ordered
correlation functions, but rather to regard them as elements of a
free-product algebra  whose expectations are obtained by averaging with
respect to all different  times. It is plausible that the {\ssf
chaotic} features of the dynamics, if any, might leave their imprinting
on the statistical properties of the  asymptotic state.

Ultimately, the testing ground of the scheme of above should consist of
those quantum systems, typically finite dimensional, that tend to
classically  chaotic systems when the dimension of the associated
Hilbert spaces increases to  infinity~[BV, BBTV, CC, D, HKS]. This involves
also an appropriate rescaling of the time.  In this paper, we analyse a
family of toy models, namely the hyperbolic toral automorphisms
quantized as in~[BNS]. These models are infinite dimensional quantum
systems arising via a  non-commutative deformation of the algebra of
continuous functions on the  two-dimensional torus. The quantum algebra
is equipped with the tracial state and endowed with a dynamics such
that the {\ssf GNS} time-evolutor coincides with  the classical
Koopman-von Neumann unitary operator.   The absolute continuity of the
spectrum of the quantum evolution guarantees those clustering 
properties which are ruled out in the commonly studied {\ssf
classically chaotic  quantum systems} which usually have discrete
spectra and quasi-periodic  time-behaviour.

The mixing properties depend on the value of a certain deformation
parameter $\theta$ and one can distinguish different randomness
conditions that lead, via a {\ssf central limit theorem}, to a variety
of statistics of fluctuations ranging from the Gaussian distribution 
to Wigner's semi-circle law.

\medskip

\beginsection{2 Statistics of multi-time correlation functions}

We briefly resume the approach of~[ABDF]. $(\A,\Theta,\phi)$ will denote a 
discrete time dynamical system where 

\item{$\bullet$}
$\A$ is a unital C*-algebra 
\item{$\bullet$}
$\Theta=\Bigl\{\Theta_t\,|\, t\in\Ir\Bigr\}$ is a discrete dynamical group
of automorphisms of $\A$ such that $X(t):=\Theta_t(X)$ represents the operator
$X$ evolved up to time $t$
\item{$\bullet$}
$\phi$ is a $\Theta$-invariant state on $\A$: $\phi\circ\Theta=\phi$.

We shall consider multi-time correlation functions of the form
$$
  {\bf t}\mapsto \phi\Bigl(X^{(1)}(t_{\nu(1)}) X^{(2)}(t_{\nu(2)}) 
  \cdots X^{(n)}(t_{\nu(n)})\Bigr)\ , 
\eqno(2.1)  
$$
where the $X^{(j)}(t_{\nu(j)})$ are operators at times $t_{\nu(j)}$ in
$\A$, ${\bf t}=\{t_1, t_2,\ldots\}\in \Ir^{\Nl_0}$ and $\nu$ maps
$\{1,2, \ldots, n\}$ into  $\Nl_0$. Two consecutive time-indices will
always be considered different, otherwise, if, say $\nu(j)=\nu(j+1)=p$,
then we write  $X^{(j)}(t_{\nu(j)}) X^{(j+1)}(t_{\nu(j+1)})=
\Bigl(X^{(j)}X^{(j+1)}\Bigr)(t_p)$. On the other hand, we allow
$\nu(j)$ to be equal to one or more of the $\nu(\ell)$ when $\ell\neq
j\pm1$.

As outlined in the introduction, while in quantum statistical mechanics
it is commonly expected that an expression  as in~(2.1) can always be
reordered by bringing together operators  at equal times, in this paper
we would like to consider dynamical situations  where the commutation
relations between operators largely separated in time  are of almost no
use. Then, the natural algebraic structure to consider is that of a
countable  {\ssf free-product} $\A_\infty=\star_{i\in\Nl_0} \A_i$  of
copies of $\A$, which is the universal  C*-algebra generated by an
identity  element $\idty$ and by ``words'' $w= X^{(1)}_{\nu(1)}
X^{(2)}_{\nu(2)}  \cdots X^{(n)}_{\nu(n)}$ that consist of
concatenations of ``letters''  $X^{(j)}\in \A$. The subscript $\nu(j)$
in $X^{(j)}_{\nu(j)}$ refers  to which copy of $\A$ the letter
$X^{(j)}$ belongs.  Concatenation, together with simplification rules,
defines the product of  words.  More specifically, the rules for
handling words are: for $X,\,Y\in\A$,  $\lambda\in\Cx$, $j\in\Nl_0$ and
$w,\,w'$ two generic words
$$\eqalignno{
  w \idty_j w'
  &= w w' &(2.2.{\rm a}) \cr
  w \bigl(X_j+ \lambda Y_j\bigr) w'
  &= w X_j w' + \lambda w Y_j w' &(2.2.{\rm b}) \cr
  w X_j Y_j w' 
  &= w (XY)_j w'\ . &(2.2.{\rm c}) 
}$$
Notice that the product $XY$ in~(2.2.c) is not concatenation, but
rather the usual operator product in the algebra $\A$. Moreover, the
adjoint $w^*$ of a word $w= X^{(1)}_{\nu(1)} X^{(2)}_{\nu(2)}  \cdots
X^{(n)}_{\nu(n)}$ equals $\bigl(X^{(n)\ast}\bigr)_{\nu(n)}
\bigl(X^{(n-1)\ast}\bigr)_{\nu(n-1)}  \cdots
\bigl(X^{(1)\ast}\bigr)_{\nu(1)}$.

We shall refer to $\A_\infty$ as to the {\ssf asymptotic free algebra}
and  equip it with an {\ssf asymptotic state} $\phi_\infty$ as follows.
With the notation
$$
  \avg\Bigl( t\mapsto f(t)\Bigr) = \lim_{T\rightarrow\infty} {1 \over T}
  \sum_{t=0}^{T-1} f(t)\ ,
\eqno(2.3)  
$$
$\nu$ any map from $\{1,2, \ldots, n\}$ into $\Nl_0$ as above, we assume
that the multiple average
$$\eqalign{
  &\bavg\Bigl({\bf t}\mapsto\phi\Bigl(X^{(1)}(t_{\nu(1)}) 
  X^{(2)}(t_{\nu(2)}) \cdots
  X^{(n)}(t_{\nu(n)})\Bigr)\Bigr):=
  \avg\Bigl(t_n\mapsto \cdots \avg\Bigl(t_2\mapsto\cr
  &\hskip 2cm \avg\Bigl(t_1\mapsto\phi\Bigl(X^{(1)}(t_{\nu(1)}) 
  X^{(2)}(t_{\nu(2)}) \cdots
  X^{(n)}(t_{\nu(n)})\Bigr)\Bigr)\Bigr)\cdots\Bigr)
  }
$$
with $n=1,2,...$, exists. Then, we define a linear functional $\phi_\infty$ on $\A_\infty$
by linearly extending the map on {\ssf elementary words}
$X^{(1)}_{\nu(1)} X^{(2)}_{\nu(2)} \cdots X^{(n)}_{\nu(n)}$, 
$$
  \phi_\infty\Bigl(X^{(1)}_{\nu(1)} X^{(2)}_{\nu(2)} \cdots
  X^{(n)}_{\nu(n)}\Bigr):=\bavg\Bigl({\bf t}\mapsto
  \phi\Bigl(X^{(1)}(t_{\nu(1)}) 
  X^{(2)}(t_{\nu(2)}) \cdots
  X^{(n)}(t_{\nu(n)})\Bigr)\Bigr)\ .
\eqno(2.4)
$$
The linear functional $\phi_\infty$ is such that it does not depend on
the insertion of identities, namely
$$\eqalign{
  &\avg\Bigl(t_3\mapsto\avg\Bigl(t_2\mapsto\avg\Bigl(t_1\mapsto
  \phi\Bigl(\idty(t_1)\idty(t_2)X(t_3)Y(t_1)Z(t_3)\Bigr)\Bigr)\Bigr)\Bigr)
  =\cr
  &\quad\avg\Bigl(t_3\mapsto\avg\Bigl(t_1\mapsto
  \phi\Bigl(X(t_3)Y(t_1)Z(t_3)\Bigr)\Bigr)\Bigr)\ .
}$$
Thus, $\phi_\infty\Bigl(X_3Y_1Z_3\Bigr)= \phi_\infty\Bigl(X_2Y_1Z_2\Bigr)$
by relabelling the dummy summation indices.
Moreover, the linear functional $\phi_\infty$ is positive and one has
the following general result:

\noindent
{\bf Proposition~1.}
{\it 
 The functional $\phi_\infty$ defined
 in~(2.4)  extends to a state on $\A_\infty$. Furthermore, 
 $$
   \phi_\infty\circ {\bf \Theta_s}= \phi_\infty \qquad\hbox{and}\qquad
   \phi_\infty\circ \alpha_\theta= \phi_\infty\ , 
 $$
 where $\bf \Theta_s$ satisfies ${\bf \Theta_s}(X_j):=
 \bigl(\Theta_{s_j}(X)\bigr)_j$  for $j\in\Nl_0$ and $X\in\A$, whereas
 $\theta$ is any order preserving injective transformation of $\Nl_0$
 and $\alpha_\theta$ is the $\ast$-homomorphism of $\A_\infty$
 determined by $\alpha_\theta(X_j):= X_{\theta(j)}$.
}

\noindent
We shall say that the asymptotic state $\phi_\infty$ is 
{\ssf permutation invariant} when 
$$
  \phi_\infty\Bigl( X^{(1)}_{\nu(1)} X^{(2)}_{\nu(2)} \cdots 
  X^{(n)}_{\nu(n)}\Bigr)= 
  \phi_\infty \Bigl(X^{(1)}_{\pi\circ\nu(1)} X^{(2)}_{\pi\circ\nu(2)} \cdots
  X^{(n)}_{\pi\circ\nu(n)}\Bigr)\qquad\forall\, X^{(j)}\in\A\ ,
\eqno(2.5)
$$
where $\pi:\Nl_0\mapsto\Nl_0$ is any bijection.
\medskip

\beginsection{3 Clustering properties and fluctuations}
 
The clustering properties of the dynamical system
$(\A,\Theta,\phi)$ determine much of the structure of the 
asymptotic state $\phi_\infty$ and hence of the statistics on the asymptotic
free algebra $\A_\infty$ associated with $(\A,\Theta,\phi)$.

Typically, two degrees of mixing are to be distinguished in quantum
systems, called {\ssf weak} and {\ssf strong clustering}. They
correspond to
$$
  \lim_{t\to\infty} \phi(X\, Y(t)\, Z)= \phi(X\, Z)\, \phi(Y), \qquad X,\, 
  Y,\, Z\in\A\ , 
\eqno(3.1)
$$
respectively 
$$
  \lim_{t\to\infty} \phi(X\, Y(t)\, Z\, S(t)\, T)= \phi(X\, Z\, T)\, \phi(Y\,
  S), \qquad X,\, Y,\, Z,\, S,\,T\in\A\ .
\eqno(3.2)
$$
The latter property, if it holds, is equivalent to {\ssf hyper-clustering},
that is to~[NT1, ABDF]
$$
  \lim_{\inf |t_i-t_j|\to\infty}\ \phi\Bigl(X^{(1)}(t_{\nu(1)}) 
  X^{(2)}(t_{\nu(2)}) \cdots X^{(n)}(t_{\nu(n)})\Bigr)= \prod_j
  \phi\Bigl(\overrightarrow{\prod}_{\kappa\in 
  \nu^{-1}(j)} X^{(\kappa)}\Bigr)\ ,
\eqno(3.3)
$$ 
where the limit is taken in such a way that all times and the differences 
between the different ones go to infinity. 
It is hyper-clustering that allows the reordering of 
multi-time correlation functions~(2.1) with repeated times, when different times
become largely separated.
On the level of the asymptotic state $\phi_\infty$, strong (equivalently
hyper-) clustering leads to:

\noindent
{\bf Proposition~2.}
{\it 
 Let $(\A,\Theta,\phi)$ be strongly clustering. Then, the asymptotic state
 $\phi_\infty$ defined on $\A_\infty$ by~(2.4) is permutation invariant.
}

Essentially, if~(3.2) holds, then using hyper-clustering, the order in
which the single time-averages in~(2.4) are performed does not  matter
for the averages coincide with the time-limits~(3.3) of the multi-time
correlation functions. Moreover, from $\phi\circ\Theta=\phi$ it follows
that $\phi_\infty(X_j)= \phi(X)$  for any $j\in\Nl_0$ and $X\in\A$,
whence 
$$
  \phi_\infty \Bigl(X^{(1)}_{\nu(1)}  X^{(2)}_{\nu(2)}  \cdots
  X^{(n)}_{\nu(n)}\Bigr)= \prod_j \phi\Bigl( 
  \overrightarrow{\prod}_{k\in \nu^{-1}(j)} X^{(k)}\Big)\ .
\eqno(3.4)
$$ 
The structure of the expectations on $\A_\infty$ calculated with
respect to the asymptotic state $\phi_\infty$ is compatible with
imposing for $j\ne k$ the commutation relations $[\A_j, \A_k]=0$ on
$\A_\infty$ and corresponds to the usual notion of  {\ssf commutative
independence~} of random variables.

It is well-known that the notion of independence embodied in~(3.4) is
incompatible with that of {\ssf free independence}~[VDN] which asks that
the correlation functions of a state $\psi$ on a free
product $\star_j \B_j$ of C*-algebras $\B_j$ satisfy
$$
  \psi\Bigl(X^{(1)}_{j_1} X^{(2)}_{j_2} \cdots X^{(n)}_{j_n}\Bigr)=0
\eqno(3.5)
$$ 
whenever $j_k\ne j_{k+1}$ and $X^{(k)}_{j_k}$ is centred (i.e.
$\psi\Bigl(X^{(k)}_{j_k}\Bigr)=0$) for all $k$. 

The two notions of independence of above are somehow extreme.
Many other possibilities for the structure of $\phi_\infty$ may arise from
the dynamical properties of $(\A,\Theta,\phi)$ and we shall try to
expose some of them by looking at limits of the form 
$$\lim_{N\to\infty} \phi_\infty\Bigl(F_N(X^{(1)}) F_N(X^{(2)}) 
  \cdots F_N(X^{(r)})\Bigr)\ , 
\eqno(3.6)
$$
and establishing a central limit theorem for the {\ssf local fluctuations}
$F_N(X)$.

\noindent
{\bf Definition~1.}
{\it 
  Let $N$ be a natural number and $X\in\A$. A local fluctuation $F_N(X)$ is the 
  following element of $\A_\infty$
  $$F_N(X):= {1\over \sqrt N} \sum_{i=1}^N \Bigl(X_i- \phi(X)\idty
  \Bigr)\ .
  \eqno(3.7)
  $$  
}

By adapting an argument in~[SvW], it can be shown that the 
cluster condition
 $$\lim_{\inf |t_i-t_j|\to\infty} \phi\Bigl(Z^{(1)}(t_{\nu(1)}) \cdots
   Z^{(j)}(t_{\nu(j)}) Y Z^{(j+1)}(t_{\nu(j+1)})\cdots
   Z^{(n)}(t_{\nu(n)})\Bigr)= 0\ ,
 \eqno(3.8)
 $$
for all centred $Y\in\A$ and $\nu:\{1,2,\ldots,n\}\mapsto\Nl_0$, which is
stronger than weak clustering~(3.1), but
weaker than strong clustering~(3.2), is sufficient to ensure that only
moments of even order contribute to the limit joint distribution of
fluctuations.

\noindent
{\bf Proposition~3.}
{\it 
 Let us assume that~(3.8) holds in $(\A,\Theta,\phi)$.
 Then, with $\phi_\infty$ defined by~(2.4) and
 $X^{(1)}, \ldots, X^{(r)}$ in $\A$ centred observables,  
 $$\lim_{N\to\infty} \phi_\infty\Bigl(F_N(X^{(1)}) 
   F_N(X^{(2)}) \cdots F_N(X^{(r)})\Bigr)=\left\{ 
   \eqalign{
   &0 \hskip 5truecm  r=2n+1 \cr
   &{1\over n!} {\sum_\nu}^{(2)} \phi_\infty\Bigl(X^{(1)}_{\nu(1)} 
   \cdots X^{(2n)}_{\nu(2n)}\Bigr) \;\; r=2n.
  }\right.
 $$
 $\sum_\nu^{(2)}$ means that we have to sum over all partitions $\nu$
 of  $\{1,2, \ldots, 2n\}$ into pairs $\bigl((\alpha_1,\beta_1)$,
 $(\alpha_2,\beta_2)$, $\ldots, (\alpha_n,\beta_n) \bigr)$ i.e.\ we choose 
 sites $\alpha_j < \beta_j$ such that $\nu(\alpha_j)= \nu(\beta_j)= j$ 
 with $j$ running from 1 to $n$.  
}

As an immediate consequence, we also have

\noindent
{\bf Corollary~1.}
{\it 
 If the asymptotic state $\phi_\infty$ is permutation invariant in
 the sense of~(2.5), and $X^{(1)}, \ldots, X^{(r)}$ are centred 
 observables of $\A$, then
 $$\lim_{N\to\infty} \phi_\infty\Bigl(F_N(X^{(1)}) 
   F_N(X^{(2)}) \cdots F_N(X^{(r)})\Bigr)=\left\{ 
   \eqalign{
   &0 \hskip 5truecm  r=2n+1 \cr
   &{\sum_{\nu,\, {\rm ord}}}^{(2)} \phi_\infty\Bigl(X^{(1)}_{\nu(1)} 
   \cdots X^{(2n)}_{\nu(2n)}\Bigr) \;\; r=2n\ ,
   }\right.
 $$
 where ${\sum_{\nu,\, {\rm ord}}}^{(2)}$ means that the sum is   over all
 {\ssf ordered} pair partitions $\nu= \bigl( (\alpha_1,\beta_1)$,
 $(\alpha_2, \beta_2)$, $\ldots, (\alpha_n, \beta_n)\bigr)$
 of  $\{1,2, \ldots, 2n\}$, i.e.\ we choose sites 
 $\alpha_j < \beta_j$ such that $\nu(\alpha_j)= \nu(\beta_j)= j$ with 
 $j$ running from 1 to $n$ and $\alpha_1< \alpha_2< \cdots< \alpha_n$.  
}

\noindent
{\bf Corollary~2.}
{\it 
 If the dynamical system $(\A,\Theta,\phi)$ is strongly clustering, then
 the fluctuations $F_N(X)$ of observables $X\in\A$ such that
 $\phi(X)=0$ and $\phi(X^2)=\sigma^2$
 tend to Gaussian random variables with zero mean and variance
 $\sigma$.
}

A {\ssf crossing} occurs in a given a pair partition $\nu$ of
$\{1,2,\ldots,2n\}$ into $n$ pairs $(\alpha_j,\beta_j)$, with
$\alpha_j<\beta_j$, when 
$$
  \alpha_j<\alpha_k<\beta_j<\beta_k\quad\hbox{for some}\quad 
   j,k\in\{1,2,\ldots,n\}\ .
\eqno(3.10)
$$
Denoting by $c(\nu)$ the number of crossings in $\nu$, we say that
$\nu$ is non-crossing if $c(\nu)=0$. If $\nu$ is a non-crossing pair
partition of $\{1,2,\ldots,2n\}$, then its pairs $(\alpha_j,\beta_j)$,
$j=1,2,\ldots,n$, are nested, that is if $\alpha_j<\alpha_k<\beta_j$
for some $j,k$, then also $\beta_k<\beta_j$. According to expectations
associated with pair partitions, one has various types of generalized
Brownian motions~[BKS,BS,S,SvW,vLM], that can be  characterized by the
contribution $-1\leq q\leq 1$ of each crossing
$$
  \phi_\infty\Bigl(X^{(1)}_{\nu(1)} \cdots X^{(2n)}_{\nu(2n)}\Bigr)=
  q^{c(\nu)}\prod_{k=1}^n\phi\Bigl(X^{(\alpha_k)}X^{(\beta_k)}\Bigr)\ .
\eqno(3.11)
$$
For $q=\pm1$ the Brownian motion is termed {\ssf Bosonic}, respectively 
{\ssf Fermionic}, 
while for $q=0$ the {\ssf $q$-deformed Brownian motion} is called 
{\ssf Free}~[VDN]
and to it only non-crossing pair partitions do contribute.

It turns out that if, besides the cluster condition~(3.8), it also
holds that, for any time independent choice of observables $A$ and $C$ 
and centred observables $X$, $Y$, $B$, 
$$
  \avg\Bigl(t\mapsto \phi\Bigl(A X(t) B Y(t) C\Bigr)\Bigr)= 0\ ,
\eqno(3.12)
$$
then a free statistics for the fluctuations of temporal averages emerges.
Namely,

\noindent
{\bf Proposition~4.}
{\it 
 Let $(\A,\Theta,\phi)$ satisfy conditions~(3.8) and~(3.12). 
 Then, with $\phi_\infty$ the asymptotic state on $\A_\infty$ defined
 by~(2.4)
 and $X^{(1)}, \ldots, X^{(2n)}$ centred observables in $\A$,  
 $$\lim_{N\to\infty} \phi_\infty\Bigl(F_N(X^{(1)}) F_N(X^{(2)}) \cdots
 F_N(X^{(2n)})
   \Bigr)= {1\over n!}{\sum_\nu}^{(2)} \delta_{0,c(\nu)} 
    \prod_{k=1}^n\phi\Bigl(X^{(\alpha_k)}X^{(\beta_k)}\Bigr)\ ,
 $$ 
 where the sum extends over all non-crossing pair partitions $\nu= 
\bigl((\alpha_1,
 \beta_1), (\alpha_2,\beta_2),\ldots$, $(\alpha_n,\beta_n)\bigr)$ of
 $\{1,2,\ldots,2n\}$ and $c(\nu)$ denotes the number of crossings in $\nu$.
 If the asymptotic state $\phi_\infty$ is also permutation invariant, then
 $$\lim_{N\to\infty} \phi_\infty\Bigl(F_N(X^{(1)}) F_N(X^{(2)}) \cdots
 F_N(X^{(2n)})
   \Bigr)= {\sum_\nu}^{(2)} \delta_{0,c(\nu)} 
    \prod_{k=1}^n\phi\Bigl(X^{(\alpha_k)}X^{(\beta_k)}\Bigr)\ ,
 $$
 where the sum now extends over all ordered and non-crossing pair partitions 
$\nu$ of
 $\{1,2,\ldots$, $2n\}$.
}

\noindent
{\bf Corollary~3.}
{\it
 If the dynamical system $(\A,\Theta,\phi)$ satisfies conditions~(3.8),~(3.1)
 and the asymptotic state $\phi_\infty$ is permutation invariant as
 in~(2.5),
 then the fluctuations $F_N(X)$ of observables
 $X\in\A$ such that $\phi(X)=0$ and $\phi(X^2)=\sigma^2$ tend to
 semicircularly distributed random  variables with zero mean and variance
 $\sigma$.
}

\medskip

\beginsection{4 Quantized automorphisms of the torus}

\noindent
We shall consider discrete time dynamical systems $(\A,\Theta,\phi)$
constructed as follows. Let $u$ and $v$ be two unitary operators
satisfying the commutation relations
$$
  u\, v = \e^{2i\pi \theta}\ v\, u \ ,
$$ 
where $0\leq\theta< 1$ will be referred to as the {\ssf deformation
parameter}.  Then, we construct the Weyl unitaries 
$$
  W_\theta(\m) :=\e^{-i\pi\theta m_1m_2}\ u^{m_1}\, v^{m_2}\ ,
$$
indexed by two-dimensional vectors with integral components
$\m=(m_1,m_2)\in\Ir^2$. The operators $W_\theta(\m)$ satisfy the
Weyl relations
$$
  W_\theta(\m)W_\theta(\n) =
    \e^{i\pi\theta\sigma(\m,\n)}\ W_\theta(\m + \n) =
    \e^{2i\pi\theta\sigma(\m,\n)}\
  W_\theta(\n)W_\theta(\m)\ ,
\eqno(4.1) 
$$
with $\sigma$ the {\ssf symplectic form} 
$$
  \sigma(\m,\n):=m_1n_2-m_2n_1\ .
$$
Notice that $W_\theta(\m)^*= W_\theta(-\m)$ and that products
$\prod_jW_\theta(\m_j)$ are reducible to just a single
$W_\theta(\sum_j\m_j)$ multiplied by a phase. 

There exists a unique C*-norm on the linear span of the Weyl unitaries
and the C*-algebra $\A$ is the completion with respect to that norm of
$\{W_\theta(f)\mid f$ complex function on $\Ir^2$ with bounded
support$\}$, where 
$$
  W_\theta(f):=\sum_{\m\in\F}\, f(\m)\, W_\theta(\m)\ .
\eqno(4.2)
$$

The dynamics $\Theta$ is determined by the linear extension to $\A$
of the map
$$
  \Theta(W_\theta(\m)):= W_\theta(T\m)\ .
\eqno(4.3)
$$
$T=\pmatrix{a&b\cr c&d}$ is a $2\times 2$ matrix with integer entries
such that $ad-bc=1$ and $a+d>2$. In this way, the eigenvalues
$\lambda_+=\lambda> 1$ and $\lambda_-=\lambda^{-1}$ of $T$ are
irrational and the associated  eigenvectors $\v_\pm$ correspond to
stretching, respectively shrinking, directions with positive,
respectively negative, Lyapounov exponent $\pm\log\lambda$. Since
$\sigma(T\m,T\n)=\sigma(\m,\n)$, the map in~(4.3) preserves
the commutation relations~(4.1) and thus extends to an automorphism of
$\A$.

Finally, the reference state $\phi$ is chosen to be the unique tracial
state on $\A$ given by
$$
  \phi\bigl(W_\theta(\m)\bigr):= \delta_{\o,\m}\ .
\eqno(4.4)
$$
The state $\phi$ is clearly $\Theta$-invariant.

\noindent
{\bf Remark. }
Via the GNS construction an infinite-dimensional Hilbert space 
representation of $\A$ is obtained, even in the case of rational 
deformation parameters $\theta$. This makes the quantization procedure
introduced above drastically different from the ones presented in the
literature~[BBTV,D] where $\theta$ is rational, typically $1/N$ and
periodicity is imposed, namely 
$$
  u^N=e^{i\alpha}\ \idty\, ,\qquad v^N=e^{i\beta}\ \idty\ .
$$ 
In such cases, the resulting algebra acts on an $N$-dimensional Hilbert
space and the quantized hyperbolic automorphisms of the torus are a
useful testing ground for studying the classical limit of quantized
classically chaotic systems. Indeed, notice that by setting $\theta=0$
(or letting $N\to\infty$), the Weyl commutation relations~(4.1) are
implemented by the exponential functions on the torus. This means that
the algebra $\A$ becomes the Abelian C*-algebra of continuous functions
on the torus which can be used to construct the corresponding
algebraic  classical dynamical system. 
\medskip

\beginsection{4.1 Number theoretical interlude}

Contrary to the finite-dimensional quantization procedures mentioned above,
where quasi-periodicity spoils any true relaxation property, the dynamical
systems $(\A,\Theta,\phi)$ enjoy, depending on $\theta$, sufficiently strong 
clustering properties. The behaviour of
commutators of observables largely separated in time is determined by the
value of the deformation parameter
$$
  \Bigl[W_\theta(\m)\,,\,W_\theta(T^t\n)\Bigr]=
  \Bigl(1- \e^{2\pi i\theta\sigma(\m,T^t\n)}\Bigr)
  W_\theta(\m + T^t\n)\ ,\qquad t\in\Ir\ . 
$$ 
If $\theta$ is rational, $\theta\sigma(\m,T^t\n) \mod 1$ can only assume
a finite number of values, and hence no clear limiting behaviour occurs,
unless of course
$\theta=0$. Some interesting behaviour occurs for peculiar 
irrational values of the deformation parameter. We shall first examine 
such a possibility in more
detail and study the convergence $\mod 1$ of the exponent
$\theta\sigma(\m,T^t\n)$ for all $\m,\n\in\Ir^2$
when $t\to\pm\infty$.
We adapt to the present problem a technique used in~[AGL] and obtain 
as a sub-case a result derived in~[N] with other means.

\noindent
{\bf Proposition~5.}
 {\it 
 Let $T=\pmatrix{a&b\cr c&d}$ with $a,b,c,d\in \Ir$ such
 that $ad-bc=1$ and $a+d>2$.
 For any $\m=(m_1,m_2)$ and $\n=(n_1,n_2)\in \Ir^2$, define 
 $$
   \Delta(\m,\n):=(1-a)(m_1n_2+n_1m_2)- c\,m_1n_1- b\,m_2n_2\ .
 $$
 Let $\lambda>1$ and $\lambda^{-1}$ be the irrational eigenvalues of
 $T$, $\v_\pm$ the corresponding eigenvectors and set 
 $$
   \beta(t):=\Tr T^t=\lambda^{t}+\lambda^{-t}\ ,\qquad t\in\Ir\ .
   \eqno(4.5)
 $$ 
 Let $\sigma(\m,\n)$ be the symplectic form $m_1n_2- m_2n_1$ on
 $\Ir^2\times\Ir^2$ and let $\theta\in[0,1)$ be irrational,
 then, the limits
 $$
   q_\pm(\m,\n):= \lim_{t\to\pm\infty}\theta\ \sigma\bigl(\m,\
   T^t\n\bigr)\ \mod 1
 \eqno(4.6) 
 $$
 exist iff the following limit exists
 $$
   \beta:= \lim_{t\to+\infty}{\theta\over\lambda^2-1}\,\beta(t)\ \mod 1\ . 
   \eqno(4.7)
 $$
 When the limit in~(4.7) exists, its possible values are the
 rational numbers
 $$
   \beta_r= {r\over\beta(1)-2}\ ,\quad 0\le r\le \beta(1)-3\ .
 \eqno(4.8)
 $$
 In such cases, the limits~(4.6) are of the form
 $$
   q^r_\pm(\m,\n)=\pm\beta_r\,\Delta(\m,\n)\ \mod 1\ .
 $$
 }
 
 \noindent
 {\bf Proof:} 
 Let $\v_\pm$ be the normalized eigenvectors of $T$.
 They satisfy $T\v_\pm=\lambda^{\pm 1}\v_\pm$ with components 
 with respect to the basis $\{(1,0),\ (0,1)\}$ given by
 $$v_\pm(1)=b/\sqrt{b^2+(a-\lambda^{\pm1})^2}\ , \quad 
   v_\pm(2)=(\lambda^{\pm 1}-a)/\sqrt{b^2+(a-\lambda^{\pm1})^2}\ .
 $$  
 By expanding $\m\in\Ir^2$ as $\m=c_+(\m)\v_+ + c_-(\m)\v_-$ and
 using $(a-\lambda)(a-\lambda^{-1})=-bc$, we explicitly compute
 $$\eqalign{
   \sigma\bigl(\m,\ T^t\n\bigr)
   &={m_1n_2\lambda^{t+2}+m_2n_1\lambda^t-
   \lambda^{t+1}(c\,m_1n_1+b\,m_2n_2+a\,m_1n_2+a\, m_2n_1)
   \over\lambda^2-1}\cr &-{m_1n_2\lambda^{-t}+m_2n_1\lambda^{-t+2}-
   \lambda^{-t+1}(c\,m_1n_1+b\,m_2n_2+a\,m_1n_2+a\, m_2n_1)
   \over\lambda^2-1}\ .
 }$$
 Then, setting $\phi:=\theta/(\lambda^2-1)$, one works out that
 $$\eqalign{
   q_+(\m,\n)
   &=\lim_{t\to+\infty}\phi\Bigl(m_1n_2\beta(t+2)+m_2n_1\beta(t)-
   \Delta_1(\m,\n)\beta(t+1)\Bigr)\, \mod 1 \cr
   q_-(\m,\n)
   &=-\lim_{t\to-\infty}\phi\Bigl(m_1n_2\beta(-t)+m_2n_1\beta(-t+2)-
   \Delta_1(\m,\n)\beta(-t+1)\Bigr)\, \mod 1\ ,
 }$$
 where $\Delta_1(\m,\n):=a(m_1n_2+n_1m_2) + cm_1n_1 +b m_2n_2$.
 Therefore, if the limit in~(4.7) exists, then the two limits above
 exist. Vice versa, if the previous two limits exist for all
 $\m,\n\in\Ir^2$, then, choosing proper values for $\m$ and $\n$, the
 limit in~(4.7) exists as well.

 \noindent
 The traces in~(4.5) obey the recursion relations
 $$
   \beta(t)=\beta(1)\beta(t-1)\ -\ \beta(t-2)\ ,\quad t\ge 2\ ,
 $$  
 with $\beta(0)=2$ and $\beta(1)=\lambda+\lambda^{-1}$.
 Thus, if the limit in~(4.7) exists, $\beta$ is determined by 
 $$
   \beta\, \bigl(\beta(1)- 2\bigr)= 0\, \mod 1\ ,
 $$
 and ranges $\mod 1$ among the rationals in~(4.8). The explicit form of
 the limit $q_\pm(\m, \n)$ is obtained by inserting in the expressions
 for $q_\pm(\m,\n)$ the values of $\beta_r$.
 \QED
 
The previous proposition gives the explicit form of the
limits in~(4.7) if they exist. 
The next result concerns the values of $\theta$ such that this is indeed the
case. 

\noindent
 {\bf Proposition~6.}
 {\it 
 $\hbox{Lim}_t\, \beta(t)\theta/(\lambda^2-1)\ \mod 1$ exists and equals 
 $\beta_r= r/(\beta(1)-2)$, with $0\le r\le \beta(1)-3$, 
 iff
 $$
   \theta = \theta^r_\ell:= \lambda\,\ell +
   (\lambda-1)\,\beta_r\ \mod 1\,,\qquad \ell\in\Ir\ .
 \eqno(4.9)
 $$
 }

\noindent
{\bf Proof:}
 We define $\phi:=\theta/(\lambda^2-1)$ and divide the proof
 in three steps. \nl
 First, we write $\phi\,\beta(t)= a(t)+b(t)$, where $a(t)$ is the
 largest natural number smaller than $\phi\ \beta(t)$. If the limit in~(4.7)
 exists, then $\lim_t b(t)= \beta_r$ for $0\le r\le \beta(1)-3$.
 Let $\epsilon(t):= b(t)- \beta_r$ and consider the quantities
 $\psi(t):= \phi\beta(t)- \beta_r= a(t)+ \epsilon(t)$. They are easily
 showed to obey the recursion relations
 $$
   \psi(t+2)= \beta(1)\psi(t+1)- \psi(t)\ +\ r\ ,\qquad t\ge 0\ .
 $$
 The latter can be rewritten as 
 $$
   a(t+2)- \beta(1)a(t+1)+ a(t)- r= \beta(1)\epsilon(t+1)-
   \epsilon(t+2)- \epsilon(t)\ ,  
 $$
 whence we deduce that there must exist an integer $T$ such that for
 $t\ge T$ both sides of the above equality vanish. Indeed, the l.h.s.
 is an integer, while the r.h.s. goes to zero with $t\to+\infty$. 
 In particular, this argument yields
 $$
   a(T+t+2)= \beta(1)\ a(T+t+1)- a(T+t)+\ r\ ,\qquad t\ge 0\ .
 \eqno(4.10)
 $$
 We prove by induction that for $t\ge 0$,
 $$
   a(T+t+2):= \gamma(t+1)\,a(T+1)- \gamma(t)\,a(T)+ r\,\sum_{k=0}^t 
   \gamma(k)\ ,
 \eqno(4.11) 
 $$
 with $\gamma(t)$ the integers such that 
 $\gamma(0)=1$, $\gamma(1)= \beta(1)$ and
 $$
   \gamma(t+2)= \beta(1)\, \gamma(t+1)- \gamma(t)\ ,\qquad t\ge 0\ .
 \eqno(4.12)
 $$
 The case $t= 0$ is obvious. Suppose that~(4.11) holds for $0\le t\le s-1$
 and rewrite~(4.10) with $t= s$ as
 $$\eqalign{
   a(T+s+2)&=a(T+1)\ \bigl(\beta(1)\gamma(s)-\gamma(s-1)\bigr)- 
   a(T)\ \bigl[\beta(1)\gamma(s-1)- \gamma(s-2)\bigr]\cr
   &+r\ \Bigl(1+\beta(1)+ \sum_{k=0}^{s-2}
   \bigl(\beta(1)\gamma(k+1)- \gamma(k)\bigr)\Bigr)\ .}
 $$  
 Using~(4.12), $a(T+s+2)$ turns out to be of the form~(4.11) with $t= s$. 
 
 \noindent
 The second step consists in observing that the coefficients
 $$
   c(t):= {\lambda^{t+2}- \lambda^{-t}\over\lambda^2-1} 
 $$
 of the expansion of $(z^2\ -\ \beta(1)\ z\ +\ 1)^{-1}$ around $z=0$
 fulfil the same recursion relations as the $\gamma(t)$ in~(4.12), with 
 the same
 initial conditions. Therefore, $\gamma(t)= c(t)$ for all $t\ge0$ and
 one has 
 $$\eqalignno{
   \beta(t)
   &=(\lambda^2-1)\ \gamma(t-2)+\lambda^{-t}\ {1+\lambda^2\over\lambda^2}
   &(4.13)\cr
   \gamma(t-1)
   &=\gamma(t-2)+1+(\beta(1)-2)\sum_{k=0}^{t-2}\gamma(k)\qquad\ t\ge 2
   &(4.14)\cr
   \lambda^{-t}
   &=\gamma(t)-\lambda\ \gamma(t-1)\ ,\hskip 3.7 truecm t\ge 1\ 
   &(4.15)\cr
   \sum_{k=0}^{t-2}\gamma(k)
   &=-{\lambda\over(\lambda^2-1)}+{\lambda^{t+1}+ \lambda^{-t+2}\over 
   (\lambda^2-1)(\lambda-1)}\ .
   &(4.16)}
 $$   
 We can use the previous relations to prove that the limit in~(4.7)
 exists and equals $\beta_r$ in~(4.8) iff 
 $$
   \theta= \lambda^{-T}\Bigl[\lambda\ a(T+1)- a(T)+
   \beta_r (\lambda-1)\Bigr]\ \mod 1\ . 
 \eqno(4.17)
 $$          
 We write $\phi= \theta/(\lambda^2-1)$ and $\phi \beta(T+t)= a(T+t)+ b(t)$
 as in the beginning of this proof.
 Then, dividing by $\beta(T+t)$ and letting
 $t\to +\infty$, one recovers~(4.17) by means of~(4.5), (4.11), (4.13) 
 and~(4.16). Vice
 versa, if $\theta$ is as in~(4.17),~(4.5) and~(4.13) yield 
 $$\eqalign{
   &\lim_{t\to+\infty} \beta(t)\ {\lambda^{-T}\over \lambda^2-1} \ 
   \Bigl(\lambda\ a(T+1)-a(T)\Bigl)\, \mod 1= 0, \cr
   &\lim_{t\to+\infty}\beta(t)\ \lambda^{-T}\ \beta_r (\lambda-1)\ \mod 1= 
   \lim_{t\to+\infty} \beta_r\bigl(\beta(t+1)- \beta(t)\bigr)\ \mod 1 \cr 
   &\phantom{\lim_{t\to+\infty}\beta(t)\ \lambda^{-T}\ \beta_r (\lambda-1)\ 
   \mod 1}= \lim_{t\to+\infty} \beta_r\bigl(\gamma(t-1)- \gamma(t-2)\bigr)\ 
   \mod 1\ , 
  }$$
 so sufficiency follows from~(4.14), since $\beta_r= r/(\beta(1)-2)$.

 \noindent
 In the third and last step we reduce the expression~(4.17) to the
 simpler form~(4.9) by using the fact that, in~(4.17),
 the integers  $T$, $a(T+1)$ and $a(T)$ are so far undetermined.
 In fact, the sufficiency of~(4.9) to
 guarantee the existence of the limit~(4.7) is proved along the same
 lines as for that of~(4.17) above. For the necessity, we have to show
 that~(4.17) reduces to~(4.9). Using~(4.15),~(4.17) reads 
 $$
   \theta= \lambda\ d+ \beta_r (\lambda-1) \Bigl(\gamma(T)-\lambda\
   \gamma(T-1) \Bigr)\ \mod 1 \ ,
 $$
 where $d:= \gamma(T-1)\ a(T)-\gamma(T-2)\ a(1+T)$. By using~(4.14) and 
 the fact that 
 $$
   \beta_r= {r\over\beta(1)-2}= r\, {\lambda\over(\lambda-1)^2}\ ,
   \quad{\rm for}\quad 0\le r\le \beta(1)-3\ ,
 $$
 we can rewrite
 $$
   \beta_r\ (\lambda-1) \Bigr(\gamma(T)-\lambda\ \gamma(T-1)\Bigr)
   =\beta_r (\lambda-1) + d_1+\lambda\, \delta \ ,
 $$
 where $d_1:= d+r\ \sum_{p=0}^{T-2}\gamma(p)$ and  $\delta:= -r
 \sum_{p=0}^{T-1}\gamma(p)$. Since $d_1$ and $\delta$ are generic
 integers, the expression~(4.9) is obtained with $\ell= d+ d_1$.
 \QED
    
The following and last number theoretical result 
is a simple characterization of certain powers of the evolution matrix $T$.

\noindent
{\bf Proposition~7.}
{\it
 Let $T$ be a $2\times 2$ hyperbolic matrix with integer entries, 
 unit determinant and trace $\beta(1)>2$ as in Propositions~5 and~6.
 Then, the matrix $T^{\beta(1)-2}$ is congruent to the identity matrix
 $\mod (\beta(1)-2)$, in the sense that its diagonal entries
 are congruent to $1$ $\mod (\beta(1)-2)$ and the off-diagonal ones are
 congruent
 to $0$ $\mod(\beta(1)-2)$.
} 

\noindent
{\bf Proof:}
 The eigenvalues $\lambda$ and $\lambda^{-1}$ of $T$ solve
 $z^2-\beta(1)z+1=0$, therefore they satisfy the matricial
 recursion relations
 $$
   \pmatrix{\lambda^n&0\cr0&\lambda^{-n}}= 
   \beta(1)\pmatrix{\lambda^{n-1}&0\cr0&\lambda^{-(n-1)}}- 
   \pmatrix{\lambda^{n-2}&0\cr0&\lambda^{-(n-2)}}\quad\hbox{for}\quad
   n\ge3\ . 
 $$ 
 Because the eigenvalues of $T$ are non-degenerate we can find a similarity
 transformation which allows us to diagonalize $T$, such that the above
 equality may be turned into
 $$T^n=\beta(1)\, T^{n-1}- T^{n-2}\ .
 $$ 
 Then, defining $S_n:=T^n-\idty$
 $$S_n=\beta(1)\, S_{n-1}- S_{n-2}\,+ \beta(1)-2\ ,\quad
   \hbox{with}\quad S_0=0\quad\hbox{and}\quad S_1=T-\idty\ ,
 $$
 the differences $S_k-S_{k-1}$, $k\ge 2$, are connected by the following
 relation
 $$S_k-S_{k-1}=S_{k-1}-S_{k-2}+\bigl(\beta(1)-2\bigr)\bigl(S_{k-1}+1\bigr)\ ,
 $$
 whence, by telescopic summation,
 $$S_n=\bigl(\beta(1)-2\bigr)\bigl(\sum_{k=2}^n S_{k-1}+ n-1\bigr)+ S_{n-1}
   +S_1\ .
 $$
 Since all matrices $S_n$ have integral entries,
 $S_n\equiv S_1+S_{n-1}$ $\mod(\beta(1)-2)$ and, by iterating $n$ times 
 the previous congruence,
 $$ S_n\equiv n\, S_1\, \mod(\beta(1)-2)\ .
 $$
 Thus, when $n=\beta(1)-2$, $S_n$ turns out to be congruent to $0$ 
 $\mod(\beta(1)-2)$  and the result follows.
 \QED

\beginsection{4.2 Clustering properties and statistics of fluctuations}

We return now to the discussion of the ergodic properties of the
quantized hyperbolic automorphisms of the torus $(\A,\Theta,\phi)$ and
of their dependence on the deformation parameter $\theta\in[0,1)$.

\noindent
 {\bf Proposition~8.}
 {\it 
 For all $\theta\in[0,1)$, the quantized hyperbolic automorphisms 
 of the torus $(\A,\Theta,\phi)$
 are weakly clustering in the sense of~(3.1). 
 They are strongly clustering in the sense of~(3.2) iff, according to
 Proposition~5,
 $$
   \lim_{t\to+\infty} {\theta\over\lambda^2-1}\beta(t)= 0\ \mod1\ ,
 $$
 that is, according to Proposition~6, iff $\theta= \lambda\ell\, \mod 1$,
 with $\ell\in\Ir$.
 }

\noindent
 {\bf Proof:} 
 The state $\phi$ is tracial and the operators $W_\theta(f)$ in~(4.2), 
 with finitely supported $f$ on $\Ir^2$, are uniformly dense in $\A$.
 Therefore, by linearity, the cluster property in~(3.1) is proved 
 by showing that
 $$\eqalign{
   &\lim_{t\to\infty} \phi\Bigl(W_\theta(\m) W_\theta(T^t\n) 
   W_\theta(\p)\Bigr)=\cr
   &\quad \lim_{t\to\infty} \e^{\pi i\theta\bigl(\sigma(\m,T^t\n)+
   \sigma(\m+T^t\n,\p)\bigr)}\, \delta_{\o,\m + T^t\n + \p}= 0\ ,
  }$$ 
 for generic integer vectors $\m,\, \n$ and $\p\in\Ir^2$. The
 commutation relations~(4.1) and the definition~(4.4) of the state
 $\phi$ have been used to derive the equality of above. The limit is
 equal to zero because of the hyperbolic character of the matrix $T$.
 Indeed, by means of the eigenvectors $\v_\pm$, one puts into evidence
 the Lyapounov exponent $\log\lambda$,
 $$ 
   T^t\n= c_+(\n)\lambda^t\v_+ + c_-(\n)\lambda^{-t}\v_-\ .
 $$
 It is thus evident that in the limit of large $|t|$, the condition
 $\m+T^t\n+\p=0$ cannot be fulfilled.
 
 \noindent
 Analogously, one calculates
 $$\phi\Bigl(\Bigl[W_\theta(\m)\,,\,W_\theta(T^t\n)\Bigr]^*
   \Bigl[W_\theta(\m)\,,\,W_\theta(T^t\n)\Bigr]\Bigr)=
   4\sin^2\bigl(\pi\theta\sigma(\m,T^t\n)\bigr)\ . 
 $$ 
 Strong clustering~(3.2) implies that the right hand side of  the above
 equation must vanish, that is that $\lim_t \theta\, \sigma(\m,T^t\n)\,
 \mod 1= 0$, for all $\m$ and $\n\in\Ir^2$.  On the other hand, if
 $\lim_t\theta\sigma(\m, T^t\n)\, \mod 1$ equals $0$  for all integer
 vectors $\m$ and $\n$, then all commutators as the  one above vanish
 in norm and therefore the dynamical system is even norm-asymptotic
 Abelian and hence, as it is weakly clustering, it is also strongly
 clustering.
 \QED

Next we shall compute the asymptotic state $\phi_\infty$ and the
fluctuations for  the hyperbolic toral automorphisms $(\A,\Theta,\phi)$
with deformation parameter $\theta$. We shall see that, depending on
the choice of the deformation parameter, these aspects can be quite
different. We shall rely on
formula~(2.4) with single-time averages as in~(2.3). The ``words'' in the
asymptotic free algebra consist now of free products of elements of the
type $W_\theta(f^{(j)})_{\nu(j)}$ with $W_\theta(f)$ given by~(4.2). Because of
linearity of $f\mapsto W_\theta(f)$ we may restrict ourselves to studying
multi-time correlation functions of Weyl operators
$$
  {\bf t}\mapsto\phi\Bigl(W_\theta\bigl(T^{t_{\nu(1)}}\n^{(1)}\bigr)\,
  W_\theta\bigl(T^{t_{\nu(2)}}\n^{(2)}\bigr)\cdots
  W_\theta\bigl(T^{t_{\nu(n)}}\n^{(n)}\bigr)\Bigr) \ ,
$$
where $\nu$ maps $\{1,2,\ldots,n\}$ into $\{1,2,\ldots,s\}$.

The following lemma can be found in~[NT2]. For the sake of completeness we
provide the proof as well.

\noindent
{\bf Lemma~1.}
{\it
 Let $T$ be any hyperbolic matrix defining the dynamics of
 $(\A,\Theta,\phi)$ as in~(4.3) and let $\sigma$ be the symplectic form
 $\sigma(\m,\n)=m_1n_2- m_2n_1$ with $\m,\, \n\in\Ir^2$. For almost all
 $\theta\in[0,1)$, the orbits $k\in\Ir\mapsto \theta\sigma(\m,T^k\n)$
 are uniformly distributed over the circle $\Ts$ of perimeter 1.
}

\noindent
{\bf Proof:}
 Uniform distribution means that $\forall f\in C(\Ts)$ we get
 $$
   \lim_{N\to\infty} {1\over N} \sum_{k=1}^{N} f(\theta\sigma(\m, T^k\n))=
   \int_0^1 dx\, f(x)\ .
 $$
 This is equivalent to requiring that 
 $$\overline{f}_N(\theta):= {1\over N} \sum_{k=1}^N \e^{2\pi
   in\theta\sigma(\m, T^k\n)}\longrightarrow 0 \qquad \forall n\in\Ir_0 \ .
 $$
 To see whether this holds for almost all $\theta$, we consider
 $$\int_0^1 d\theta\, |\overline{f}_N(\theta)|^2= {1\over N^2}
   \sum_{k,k'=1}^N \int_0^1 d\theta\, \e^{2\pi in\theta \sigma(\m,
   (T^k- T^{k'})\n)}\ .
 $$
 Because $\sigma\bigl(\m, (T^k-T^{k'})\n\bigr)$ is an integer, we have that
 $$\int_0^1 d\theta\, \e^{2\pi in\theta
   \sigma(\m,(T^k-T^{k'})\n)}=
   \cases{1   &if $\sigma\bigl(\m,(T^k- T^{k'}\bigr)\n)= 0$ \cr
          0   &else}\ .
 $$	          
 And, because $\sigma\bigl(\m, (T^k- T^{k'})\n\bigr)= 0\ \forall\
 \m\in  \Ir^2   \Leftrightarrow T^k\n= T^{k'}\n\Leftrightarrow k=k'$,
 we have that
 $$
   \int_0^1 d\theta\, |\overline f_N(\theta)|^2= 1/N^2\sum_{k=k'=1}^N
   1= 1/N\longrightarrow 0 \hbox{ for } N\to\infty \ .
 \eqno{{\vrule height 1.2ex width 1.2ex }\quad} 
 $$ 

\noindent
Using the previous lemma and Propositions~5 and~6, we deduce

\noindent
{\bf Corollary~4.}
{\it 
 $a)\ $
 There exists a set $\Z\subset [0,1)$ of measure 1 such that for all $\theta\in
 \Z$ and for all
 $\m,\,\n\in\Ir^2$ ($\neq \o$), we have  
 $$\lim_{T\to\infty}{1\over T}\sum_{t=1}^{T}\e^{2\pi i\theta\sigma(\m,T^t\n)}=0
 \ . 
 $$ 
 $b)\ $ 
 If  $\lim_t\theta\sigma(\m,T^t\n)\, \mod 1$ exists, then
 $$\lim_{T\to\infty}{1\over T}\sum_{t=1}^{T}\e^{2\pi i\theta\sigma(\m,T^t\n)} 
 =\e^{2\pi i\beta_r\Delta(\m,\n)}\ ,
 $$
 where $\beta_r$ and $\Delta(\m,\n)$ are defined in Proposition~5.
}

 In order to proceed with the construction of an asymptotic state on $\A_\infty$
 following the prescription of~(2.4), we need a preliminary technical result
 which is proved in~[ABDF].
 
 \noindent
{\bf Lemma~2.}
{\it
  For $d,k\in \Nl$ define
  $$\Delta_d^k (t_1,\ldots,t_k) = \cases{
   0 & if $|t_i - t_j|\leq d$ for some $1\leq i \not= j \leq k$ \cr
   1 & else}\ . 
  $$
  Then, if the multiple average~(2.14) of a uniformly bounded function 
  $f:\Nl^k\to \Cx$ exists, we have 
  $$\eqalign{
    &\avg\Bigl( t_k\mapsto \cdots \avg\Bigl( t_2\mapsto
    \avg\Bigl( t_1\mapsto f(t_1, t_2, \ldots, t_k)\Bigr)\Bigr)\cdots
    \Bigr)= \cr
    &\avg\Bigl( t_k\mapsto \cdots \avg\Bigl( t_2\mapsto
    \avg\Bigl( t_1\mapsto \Delta_d^k(t_1,\ldots,t_k) f(t_1, t_2,
    \ldots, t_k)\Bigr)\Bigr)\cdots \Bigr) \ .
  }$$ 
}

Proposition~9 and the subsequent corollary deal with the asymptotic structure
emerging from the first possibility of corollary~4, while proposition~10 and its
corollary deal with the second case. 

\noindent
{\bf Proposition~9.}
{\it 
  Let us consider a hyperbolic toral automorphism determined by a
  deformation parameter $\theta\in\Z$ as in
  corollary~4, part a). The multiple average in~(2.4), with single-time averages
  as in~(2.3), exists and the state $\phi_\infty$ it defines on $\A_\infty$
  is permutation invariant as in~(2.5). 
} 

\noindent
{\bf Proof:} 
 Because of linearity, we need only to consider expectations of words of
 the type 
 $$
   \phi_\infty\Bigl( W_\theta\bigl(\n^{(1)}\bigr)_{\nu(1)}\,
   W_\theta\bigl(\n^{(2)}\bigr)_{\nu(2)}\cdots
   W_\theta\bigl(\n^{(n)}\bigr)_{\nu(n)}\Bigr) \ ,
 $$
 which are to be computed as the multiple averages
 $$\eqalign{
   &\avg\Bigl( t_s\mapsto \cdots \avg\Bigl( t_2\mapsto \avg\Bigl( t_1\mapsto\cr
   &\hskip 2cm  \phi\Bigl(W_\theta\bigl(T^{t_{\nu(1)}}\n^{(1)}\bigr)\,
   W_\theta\bigl(T^{t_{\nu(2)}}\n^{(2)}\bigr)\cdots
   W_\theta\bigl(T^{t_{\nu(n)}}\n^{(n)}\bigr)\Bigr)\Bigr)\Bigr)\cdots\Bigr)\ . 
 }$$
 $\nu$ maps $\{1,2,\ldots,n\}$ into $\{1,2,\ldots,s\}$ and we may
 assume that all Weyl operators are different from the identity, that
 is that none of the $\n^{(j)}=0$.\nl
 Before taking the average we group all
 Weyl operators belonging to a same time, gathering hereby a phase
 factor according to the commutation relations~(4.1). Then, using
 Lemma~2, we reduce the problem to the computation of the multiple 
 time-average of the phase factors. By the assumption on the deformation
 parameter $\theta$, these averages are always zero unless they are
 time-independent. This proves the permutation invariance 
 of $\phi_\infty$.
 
 \noindent 
 With $p\in\{1,2,\ldots,s\}$, let $I_p$ denote the set of natural
 numbers  
 $i\in\{1,2,\ldots,n\}$ such that $\nu(i)=p$. 
 For any given $p\in\{1,2,\ldots,s\}$, the Weyl operators
 $W_\theta\bigl(T^{t_{\nu(i)}}\n^{(i)}\bigr)$, with $i\in I_p$, 
 can be brought together, from right to left, to form the composite word
 $$W_{I_p}(t_p):= W_\theta\bigl(T^{t_p}\n^{(i_1)}\bigr)\, 
   W_\theta\bigl(T^{t_p}\n^{(i_2)}\bigr)\cdots
   W_\theta\bigl(T^{t_p}\n^{(i_{c_p})}\bigr)\ . 
 $$
 Using~(4.1), the regrouping produces a phase factor
 $$\exp\biggl( \sum_{i\in I_p} \sum_{k\in K(i)} \sum_{j\in J_i(k)} 
   2\pi i\theta \sigma\Bigl(T^{t_k}\n^{(j)},\, T^{t_p} \n^{(i)}\Bigr)\biggr) \ .
 $$ 
 The index set $K(i)$ specifies those times $t_k\neq t_p$
 that are encountered while  commuting
 $W_\theta\bigl(T^{t_p}\n^{(i)}\bigr)$, $i\in I_p$, over the
 various  $W_\theta\bigl(T^{t_{\nu(i)}} \n^{(i)}\bigr)$ in order to
 concatenate the former word to the right of a previous word
 $W_\theta\bigl(T^{t_p}\n^{(j)}\bigr)$, $j\in I_p$.
 
 \noindent 
 We shall regroup the Weyl operators by first bringing together all those at 
 time $t_1$, then all those at time $t_2$ and so on. 
 We end up with 
 $$\eqalignno{
   &\phi\Bigl(W_\theta\bigl(T^{t_{\nu(1)}}\n^{(1)}\bigr)\, 
   W_\theta\bigl(T^{t_{\nu(2)}}\n^{(2)}\bigr)\cdots
   W_\theta\bigl(T^{t_{\nu(n)}}\n^{(n)}\bigr)\Bigr)= \cr 
   &\hskip 2cm F\Bigl(t_1,t_2,\ldots,t_s\Bigr)\,
   \phi\Bigl(W_{I_1}(t_1)\, W_{I_2}(t_2)\, \cdots W_{I_s}(t_s)\Bigr)\ , 
   &(4.18)  
 }$$ 
 where
 $$\eqalign{F\Bigl(t_1,t_2,\ldots,t_s\Bigr) 
   &=\ \prod_{p=1}^s\  F_p(t_p,t_{p+1},\ldots,t_s)\qquad{\rm with}\cr 
   F_p(t_p,t_{p+1},\ldots,t_s) 
   &=\exp\biggl( \sum_{i\in I_p} \sum_{k\in K_{p+1}(i)}
   \sum_{j\in J_i(k)} \, 2\pi i\theta \sigma\Bigl(T^{t_k} \n^{(j)},\, 
   T^{t_p}\n^{(i)}\Bigr)\biggr)\ , 
 }
 \eqno(4.19)$$
 and $K_{p+1}(i)$ labels times $t_k$ with $k\ge p+1$.
 Next, using Lemma~2, we can always restrict ourselves in the multiple
 time-averages to times $t_{\nu(j)}$ that are sufficiently separated from 
 one another. 
 As the dynamics $\n\mapsto T \n$ is hyperbolic on $\Ir^2$, the
 separation between different times can always be chosen in such a way that 
 $$\eqalignno{ 
   &\nu(i)\ne \nu(j)\Longrightarrow T^{t_{\nu(i)}}\n^{(i)}+ T^{t_{\nu(j)}} 
   \n^{(j)}\ne \o 
   \qquad\hbox{and} \cr 
   &\sum_{j=1}^n T^{t_{\nu(j)}} \n^{(j)}= \o \Longleftrightarrow \n_p:= 
   \sum_{i\in I_p} \n^{(i)}= \o \qquad\forall p=1,2,\ldots, s \ . 
   &(4.20)
 }$$ 
 {From}~(4.20) and~(4.4), we see that the only possibly non-vanishing averages
 are those for which $\n_p= \o$ for each $p$ separately, in which case
 $W_{I_p}=G_p\, W_\theta(\n_p)=G_p\,\idty$, $G_p$ being a suitable phase 
 obtained through~(4.1). 
 We can therefore write  
 $$\eqalignno{
   &\phi_\infty\Bigl( W_\theta\bigl(\n^{(1)}\bigr)_{\nu(1)}\, 
   W_\theta\bigl(\n^{(2)}\bigr)_{\nu(2)}\cdots
   W_\theta\bigl(\n^{(n)}\bigr)_{\nu(n)}\Bigr)= \cr 
   &\Bigl(\prod_{q=1}^s G_q\delta_{\o, \n_q}\Bigl)\, \avg\Bigl( t_s\mapsto
   \cdots \avg\Bigl( t_2\mapsto \avg\Bigl( t_1\mapsto 
   F\Bigl(t_1,t_2,\ldots,t_s\Bigr)\Bigr)\Bigr)\cdots \Bigr)\ .  
   &(4.21) 
 }$$   
 We now take the average with respect to $t_1$. In the product~(3.19), only 
 the factor $F_1$ can depend on $t_1$. 
 Since $\theta$ belongs to $\Z$, this average is either zero or
 one. Moreover, it can only be one if $F_1$ does not depend on $t_1$. 
 We can now successively average over the consecutive times $t_2, t_3, \ldots$ 
 and conclude that either the average is zero or that it is independent
 of all $t_1, t_2, \ldots, t_s$. 
 In both cases, $\phi_\infty$ is permutation invariant in the sense 
 of~(2.5).  
\QED 

\noindent
{\bf Corollary~5.}
{\it
  A hyperbolic toral automorphism determined by a deformation parameter
  $\theta\in\Z$ as in Corollary~4, part $a)$, satisfies condition~(3.8) 
  and~(3.12).
  Therefore, the fluctuations of a centred
  self-adjoint element $W_\theta(f)$ in~(4.2), are semicircularly
  distributed. 
}

\noindent
{\bf Proof:}
 Condition~(3.8) is satisfied because of the hyperbolic character of the matrix
 $T$ which implements the dynamics.
 Thus, because of linearity and since the reference state is the 
 trace, it suffices to check condition~(3.12) in the form
 $$
   \avg\Bigl(t\mapsto \phi\Bigl(W_\theta(\p)W_\theta(T^t \m)
   W_\theta(\n)W_\theta(-T^t \m) \Bigr)\Bigr)= 0 \ ,
 \eqno(4.22)
 $$
 for all $\m,\,\n,\,\p\in\Ir^2$.
 In fact, \nl
 $a)\,$ products of Weyl operators can be reduced to a single Weyl operator 
   multiplied by a phase factor; \nl
 $b)\,$ Weyl operators on the right of $W_\theta(-T^t \m)$ 
   can be moved to the left of $W_\theta(T^t \m)$ by the tracial properties of 
   the state $\phi$ and \nl
 $c)\,$ the two and only two Weyl operators at time $t$ carry opposite
   integral vectors because, otherwise, the hyperbolic character of the matrix 
   $T$ and~(4.4) would force the expectation $\phi$, and thus also its average,
   to vanish asymptotically. \nl
 Then~(4.22) is obviously true, as
 $$\eqalign{
   &\avg\Bigl(t\mapsto \phi\Bigl(W_\theta(\p)W_\theta(T^t \m)
   W_\theta(-\n)W_\theta(-T^t \m) \Bigr)\Bigr)=\cr 
   &\hskip 2cm \phi\Bigl(W_\theta(\p) W_\theta(-\n)\Bigr)\, 
   \avg\Bigl(t\mapsto \exp\Bigl(2\pi i\theta\sigma(T^t \m,\n)\Bigr)\Bigr)\ .
 }$$
 On the basis of this result, we want to compute  $\lim_N
 \phi_\infty\Bigl(F_N(W_\theta(f))^{2n}\Bigr)$ where $F_N$ is the local
 fluctuation defined in~(3.7). Notice that $W_\theta(f)$ is
 self-adjoint iff, for any $\n\in\Ir^2$,  $\overline{f(\n)}= f(-\n)$
 and it is centred iff $f(\o)=0$.  We shall denote by $\F$ the support
 of $f$. Since $\phi_\infty$ is permutation invariant, we use
 Corollary~3 and  obtain a semicircular distribution for $W_\theta(f)$
 with variance 
 $$
   \sigma=\phi(W_\theta(f)^2)= \|f\|_2^2=\sum_{\q\in\F} |f(\q)|^2\ .
 $$ 
\QED  
 
\noindent
{\bf Remark $\,$}
The previous proof could suggest that the state $\phi_\infty$ which was
constructed in Proposition~9 is a free product of traces, namely that
$$
  \phi_\infty\Bigl(X^{(1)}_{j_1} X^{(2)}_{j_2} \cdots X^{(n)}_{j_n}\Bigr)=0
$$ 
whenever $\phi\Bigl(X^{(k)}_{j_k}\Bigr)=0$ and $j_k\ne j_{k+1}$ for all
$k$. This is, however, not the case. An easy counterexample is
obtained by considering a correlation function that is independent of a
time appearing in the product of observables such as
$$
  t\mapsto \phi\Bigl(W_\theta(\n) W_\theta(T^t\n) W_\theta(-2\n)
  W_\theta(T^t\n) W_\theta(\n) W_\theta(-2 T^t\n)\Bigr)\qquad
  \n\ne\o\ .
$$
Each of the observables in the correlation function is centred but,
instead of vanishing, 
$$
  \phi_\infty\Bigl(W_\theta(\n)_1 W_\theta(\n)_2 W_\theta(-2\n)_1
  W_\theta(\n)_2 W_\theta(\n)_1 W_\theta(-2\n)_2\Bigr)= 1\ .
$$

\medskip

We shall now see that, when the deformation parameter $\theta$ is
chosen not to belong to the dense set $\Z$, but, instead, equals any of
the special values for which the second part of Corollary~4 holds, then
the multiple time-average~(2.4) still defines an asymptotic state
$\phi_\infty$, but, except when $\beta_r=0$, it is not permutation
invariant.

\noindent
{\bf Proposition~10.}
{\it 
 If the conditions of Corollary~4, part $b)$, are fulfilled, then 
 the multiple average~(2.4) exists and it defines an asymptotic
 state $\phi_\infty$ on the asymptotic free algebra $\A_\infty$.
} 

\noindent
{\bf Proof:}
 We can follow the proof of Proposition~9.
 The asymptotic state $\phi_\infty$ is determined by the expectations
 $$
   \phi_\infty\Bigl( W_\theta\bigl(\n^{(1)}\bigr)_{\nu(1)}\,
   W_\theta\bigl(\n^{(2)}\bigr)_{\nu(2)}\cdots
   W_\theta\bigl(\n^{(n)}\bigr)_{\nu(n)}\Bigr) \ ,
 $$
 where $\nu$ maps $\{1,2,\ldots,n\}$ into $\{1,2,\ldots,s\}$, $s\le n$, and
 the vectors $\n^{(j)}\in\Ir^2$ may be supposed $\neq\o$. 
 By  suitable regrouping as in formula~(4.18), the expectations above are 
 well-defined
 if the multiple time-average~(4.21) exists.
 Thus, we start considering the time $t_1$.
 By virtue of the assumption on the existence of the limits of the exponents, 
 we get
 $$\eqalignno{
  &\avg\Bigl( t_1\mapsto F\Bigl(t_1,t_2,\ldots,t_s\Bigr)\Bigr)= \cr
  &\prod_{p=2}^s F_p\Bigl(t_p, \ldots, t_s\Bigr)\avg\biggl(
  t_1\mapsto \exp\biggl( \sum_{i\in I_1}
  \sum_{k\in K_2(i)} \sum_{j\in J_i(k)} 2\pi i\theta \sigma\Bigl(T^{t_k}
  \n^{(j)},\, T^{t_1} \n^{(i)}\Bigr)\biggr)\biggr)= \cr 
  &\prod_{p=2}^s F_p\Bigl(t_p, \ldots, t_s\Bigr)\ \exp\biggl(
  \sum_{i\in I_1} \sum_{k\in K_2(i)} \sum_{j\in J_i(k)} 2\pi i\beta_r   
  \Delta\Bigl(T^{t_k} \n^{(j)},\,\n^{(i)}\Bigr)\biggr)\ .  
  &(4.23)
 }$$
 In the last equality, we have used the quantities $\beta_r$ and 
 $\Delta(\m,\n)$ which have been introduced in Propositions~5 and~6.
 Furthermore, the index sets $K_\ell(i)$, $i\in I_q$ and $\ell\ge q+1$
 contain $k\ge \ell$. 
 
 \noindent
 When  averaging with respect to $t_2$, the factor $F_2(t_2,t_3,\ldots,t_s)$ 
 tends asymptotically to a term similar to the
 exponential in formula~(4.23) of above, without $t_2$ dependence. 
 Therefore, we only have to compute the average with respect to $t_2$ of a
 contribution of the form
 $$\exp\biggl( \sum_{i\in I_1} \sum_{j\in J_i(2)} 2\pi i\beta_r\Delta
 \Bigl(T^{t_2}\n^{(j)},\,\n^{(i)}\Bigr)\biggr)
 $$    
 which may come from the presence of the time $t_2$ among those indexed by
 $k\in K_2(i)$ in the exponent in~(4.23).
 
 \noindent
 We rewrite the quantity $\Delta(\m,\n)$ introduced in 
 Proposition~5 as the scalar product
 $\Delta(\m,\n)=\langle\m,S\n\rangle$ where $S=\pmatrix{-c&1-a\cr1-a&-b}$.
 Since powers of $T$ transform $\Ir^2$ into $\Ir^2$ and $S$ has integral 
 entries, we use Proposition~7 to deduce that
 $$\Delta\Bigl(T^{m(\beta(1)- 2)+ s}\n^{(j)},\,\n^{(i)}
   \Bigr)=\langle T^s\n^{(j)},\,S\n^{(i)}\rangle\, + 
   (\beta(1)-2)\, N(\n^{(i)},\n^{(j)})\ ,
 $$
 for all $m\in \Ir$, with $0\le s\le \beta(1)-3$ and $ N(\n^{(i)},\n^{(j)})$ a 
 suitable integer. Moreover, $\beta_r=r/(\beta(1)-2)$ with $0\le 
r\le\beta(1)-3$, thus
 the following cyclic properties hold
 $$\eqalign{
   &\exp\biggl(\sum_{i\in I_1}\sum_{j\in J_i(2)} 2\pi i\beta_r
   \Delta\Bigl(T^{m(\beta(1)- 2)+ s}\n^{(j)},\,\n^{(i)}\Bigr)\biggr)=\cr
   &\qquad\exp\biggl(\sum_{i\in I_1}\sum_{j\in J_i(2)} 2\pi i\beta_r 
   \Delta\Bigl(T^s\n^{(j)},\,\n^{(i)}\Bigr)\biggr)\ .}
 \eqno(4.24)
 $$
 After the first two averages, we remain with
 $$\eqalign{
   &\avg\Bigl( t_2\mapsto\avg\Bigl( t_1\mapsto 
F\Bigl(t_1,t_2,\ldots,t_s\Bigr)\Bigr)\Bigr)= \cr
   &\prod_{p=3}^s F_p\Bigl(t_p, \ldots, t_s\Bigr)\ \exp\biggl(
   \sum_{i\in I_2} \sum_{k\in K_3(i)} \sum_{j\in J_i(k)} 2\pi i\beta_r
   \Delta\Bigl(T^{t_k} \n^{(j)},\,\n^{(i)}\Bigr)\biggr)\ \times\cr
   &\quad\exp\biggl(
   \sum_{i\in I_1} \sum_{k\in K_3(i)} \sum_{j\in J_i(k)} 2\pi i\beta_r
   \Delta\Bigl(T^{t_k} \n^{(j)},\,\n^{(i)}\Bigr)\biggr)\  D_1^2
 }$$
 where, using~(4.24),
 $$\eqalign{D_1^2
   &:=\avg\Bigl(t_2\mapsto \exp\biggl( \sum_{i\in I_1}\sum_{j\in J_i(2)}2\pi 
i\beta_r
   \Delta\Bigl(T^{t_2}\n^{(j)},\,\n^{(i)}\Bigr)\biggr)\Bigr)\cr
   &={1\over\beta(1)-2}\sum_{s=0}^{\beta(1)-3}\exp\biggl(
   \sum_{i\in I_1}\sum_{j\in J_i(2)} 2\pi i\beta_r 
   \Delta\Bigl(T^s\n^{(j)},\,\n^{(i)}\Bigr)\biggr)\ .
 }$$ 
 If we go on and consider the average with respect to $t_3$, the result is
 $$\eqalign{
   &\avg\Bigl( t_3\mapsto\avg\Bigl(t_2\mapsto\avg\Bigl( t_1\mapsto 
   F\Bigl(t_1,t_2,\ldots,t_s\Bigr)\Bigr)\Bigr)\Bigr)= \cr
   &\prod_{p=4}^s F_p\Bigl(t_p, \ldots, t_s\Bigr)\ \exp\biggl(
   \sum_{i\in I_3} \sum_{k\in K_4(i)} \sum_{j\in J_i(k)} 2\pi i\beta_r
   \Delta\Bigl(T^{t_k} \n^{(j)},\,\n^{(i)}\Bigr)\biggr)\ \times\cr
   &\quad\exp\biggl(
   \sum_{i\in I_2} \sum_{k\in K_4(i)} \sum_{j\in J_i(k)} 2\pi i\beta_r
   \Delta\Bigl(T^{t_k} \n^{(j)},\,\n^{(i)}\Bigr)\biggr)\ \times\cr
   &\quad\exp\biggl(\sum_{i\in I_1} \sum_{k\in K_4(i)} \sum_{j\in J_i(k)} 
   2\pi i\beta_r\Delta\Bigl(T^{t_k} 
\n^{(j)},\,\n^{(i)}\Bigr)\biggr)\,D_{12}^3\,D_1^2\ ,
 }$$
 where
 $$\eqalign{D_{12}^3&:={1\over\beta(1)-2}\sum_{s=0}^{\beta(1)-3}\,
   \exp\biggl(2\pi i\beta_r\biggl\{\sum_{i\in I_1} \sum_{j\in J_i(3)}
   \Delta\Bigl(T^s \n^{(j)},\,\n^{(i)}\Bigr)\ +\cr
   &\hskip 3cm 
   \sum_{k\in I_2} \sum_{q\in J_k(3)}\Delta\Bigl(T^s \n^{(q)},\,\n^{(k)}
   \Bigr)\biggr\}\biggr)\ .}
 $$
 After averaging with respect to the remaining times $t_3,t_4,\ldots,t_s$, 
 we obtain a well-defined asymptotic state $\phi_\infty$ given by
 $$\eqalignno{
   &\phi_\infty\Bigl( W_\theta\bigl(\n^{(1)}\bigr)_{\nu(1)}\,
   W_\theta\bigl(\n^{(2)}\bigr)_{\nu(2)}\cdots 
W_\theta\bigl(\n^{(n)}\bigr)_{\nu(n)}\Bigr)
   =\prod_{p=2}^s\, D^p_{12\ldots p-1}
   \cr
   &\hskip 1cm \hbox{where}\cr
   &D^p_{12\ldots p-1}:={1\over\beta(1)-2}\sum_{s=0}^{\beta(1)-3}
   \exp\biggl(2\pi i\beta_r \sum_{\ell=1}^{p-1}\sum_{i\in I_\ell}
   \sum_{j\in J_i(p)} \Delta\Bigl(T^s\n^{(j)},\,\n^{(i)}\Bigr)\biggr)\ .
   &{\vrule height 1.2ex width 1.2ex }\quad 
 }$$

The asymptotic states just constructed are not permutation invariant as
in~(2.5). Indeed, when $\beta_r\neq0$,
$$\eqalign{
  &\phi_\infty\Bigl( W_\theta(\p)_3\, W_\theta(\p)_1\,
  W_\theta(\m)_3\, W_\theta(-\m)_2\, W_\theta(-\p)_1\,
  W_\theta(\m)_2\, W_\theta(-\p-\m)_3 \Bigr)\neq\cr 
  &\hskip 1cm \phi_\infty\Bigl( W_\theta(\p)_1\, W_\theta(\p)_3\,
  W_\theta(\m)_1\, W_\theta(-\m)_2\, W_\theta(-\p)_3\,
  W_\theta(\m)_2\, W_\theta(-\m-\p)_1 \Bigr)\ .
}$$
By relabelling the summation indices, as in the proof of Proposition~9, 
the inequality of above comes about if we show that, with shortened notation 
for the successive averages,
$$\eqalign{
   &\avg_3\avg_2\avg_1 \Bigl( \phi\Bigl( W_\theta\bigl(T^{t_3}\p\bigr)\, 
   W_\theta\bigl(T^{t_1}\p\bigr)\, W_\theta\bigl(T^{t_3}\m\bigr)\, 
   W_\theta\bigl(-T^{t_2}\m\bigr)\, W_\theta\bigl(-T^{t_1}\p\bigr)\cr 
   &\hskip 3cm W_\theta\bigl(T^{t_2}\m\bigr)\, W_\theta\bigl(-T^{t_3}
   (\m+\p)\bigr)\Bigr)
   \neq\cr
   &\avg_1\avg_2\avg_3 \Bigl( \phi\Bigl( W_\theta\bigl(T^{t_3}\p\bigr)\, 
   W_\theta\bigl(T^{t_1}\p\bigr)\, W_\theta\bigl(T^{t_3}\m\bigr)\cr 
   &\hskip 3cm W_\theta\bigl(-T^{t_2}\m\bigr)\, W_\theta\bigl(-T^{t_1}\p\bigr)\, 
   W_\theta\bigl(T^{t_2}\m\bigr)\, W_\theta\bigl(-T^{t_3}(\m+\p)\bigr)\Bigr)\ . 
}$$
We now bring together from right to left the words belonging to $t_1$, $t_2$ 
and $t_3$, thus obtaining
$$\eqalign{
  & W_\theta\bigl(T^{t_3}\p\bigr) 
  W_\theta\bigl(T^{t_1}\p\bigr) W_\theta\bigl(T^{t_3}\m\bigr) 
  W_\theta\bigl(-T^{t_2}\m\bigr) W_\theta\bigl(-T^{t_1}\p\bigr) 
  W_\theta\bigl(T^{t_2}\m\bigr) W_\theta\bigl(-T^{t_3}(\m+\p)\bigr)=\cr
  &\hskip 1cm\exp\biggl(2\pi i\theta\Bigl(\sigma\bigl(T^{t_2}\m,T^{t_1}\p\bigr)- 
  \sigma\bigl(T^{t_3}\m,T^{t_1}\p\bigr)\Bigr)\biggr)\,\idty\ .
}$$  
Then, arguing as in the proof of Proposition~10, we compute
$$\eqalignno{
   &\avg_3\avg_2\avg_1 \biggl(
   \exp\biggl(2\pi i\theta\Bigl(\sigma\bigl(T^{t_2}\m,T^{t_1}\p\bigr)- 
   \sigma\bigl(T^{t_3}\m,T^{t_1}\p\bigr)\Bigr)\biggr)\biggr)=\cr
   &\avg_3\avg_2\biggl(
   \exp\biggl(2\pi i\beta_r\Bigl(\Delta\bigl(T^{t_2}\m,\p\bigr)- 
   \Delta\bigl(T^{t_3}\m,\p\bigr)\Bigr)\biggr)\biggr)=\cr
   &\left({1\over\beta(1)-2}\right)^2\ \left|\, \sum_{s=0}^{\beta(1)-3}\, 
   \exp\Bigl(2\pi i\beta_r\Delta\bigl(T^s\m,\p\bigr)\Bigr)\, \right|^2\ .
   &(4.25)
}$$
On the other hand,
$$ \eqalignno{
   &\avg_1\avg_2\avg_3 \biggl(
   \exp\biggl(2\pi i\theta\Bigl(\sigma\bigl(T^{t_2}\m,T^{t_1}\p\bigr)- 
   \sigma\bigl(T^{t_3}\m,T^{t_1}\p\bigr)\Bigr)\biggr)\biggr)=\cr
   &\avg_1\avg_2\biggl(
   \exp\Bigl(2\pi i\theta\sigma\bigl(T^{t_2}\m,T^{t_1}\p\bigr)\Bigr)
   \exp\biggl(2\pi 
i\beta_r\Bigl(\Delta\bigl(T^{t_1}\p,\m\bigr)\Bigr)\biggr)\biggr)=\cr
   &\avg_1\biggl(\exp\Bigl(-2\pi i\beta_r\Delta\bigl(T^{t_1}\p,\m\bigr)\Bigr)
   \exp\Bigl(2\pi i\beta_r\Delta\bigl(T^{t_1}\p,\m\bigr)\Bigr)\biggr)\,=1\ . 
}$$
Clearly, the expression~(4.25) can be made different from $1$ by suitably 
choosing $\m$ and $\p$.    

Notice that, if $\beta_r=0$, which is the case when $r=0$, then~(4.25) 
equals $1$ as well. In fact, Proposition~8 tells us that the system 
$(\A,\Theta,\phi)$ is strongly asymptotically Abelian and Proposition~2
ensures that the asymptotic state is then automatically  permutation
invariant.

In order to characterize the statistics of fluctuations in the
asymptotic  algebra $\A_\infty$, when the asymptotic state
$\phi_\infty$ is defined as in  Proposition~10, we show that, besides
being weakly clustering, the  dynamical systems $(\A,\Theta,\phi)$
satisfy condition~(3.8).

\noindent
{\bf Proposition~11.}
{\it
 Let $(\A,\Theta,\phi)$ be a quantized toral automorphism with deformation 
 parameter $\theta$ such that $\lim_t \theta \beta(t)/(\lambda^2-1)\ 
 \mod 1= \beta_r$ as in Proposition~5, with $r\neq 0$.
 Then,  
 $$\eqalign{
   &\lim_{\inf |t_i-t_j|\to\infty} \phi\Bigl(W_\theta(f^{(1)})(t_{\nu(1)}) 
   \cdots W_\theta(f^{(j)})(t_{\nu(j)}) W_\theta(g) \cr
   &\hskip 3cm W_\theta(f^{(j+1)})(t_{\nu(j+1)})\cdots
   W_\theta(f^{(n)})(t_{\nu(n)})\Bigr)= 0 \ ,
 }
 $$
 where $W_\theta(f^{(\ell)})$, $W_\theta(g)$ are defined as in~(4.2), 
 by means of finitely supported functions, and
 $$
   W_\theta(f^{(\ell)})(t_{\nu(j)})=\sum_{\m\in\F_\ell} f^{(\ell)}(\m) 
   W_\theta\Bigl(T^{t_{\nu(\ell)}}\m\Bigr))\ .
 $$ 
}

\noindent
{\bf Proof:}
Because of linearity and the finiteness of the supports $\F_\ell$ of
$f^{(\ell)}$, we can  restrict our  considerations to expectations of the
form
 $$\eqalign{
   &\phi\biggl(W_\theta\Bigl(T^{t_{\nu(1)}}\p^{(1)}\Bigr) \cdots
   W_\theta\Bigl(T^{t_{\nu(j)}}\p^{(j)}\Bigr) W_\theta(T^s\m)\cr 
   &\hskip 3cm W_\theta\Bigl(T^{t_{\nu(j+1)}}\p^{(1j+1)}\Bigr) \cdots
   W_\theta\Bigl(T^{t_{\nu(n)}}\p^{(n)}\Bigr)\biggr)= 0\ ,
 }$$
where $\m\neq\o$, for the corresponding word is assumed to be centred.
If we now group together Weyl operators at equal times, $T^s\m$ is not
matched by any of the other operators. On the other hand, using Lemma~2
and considering the smallest difference $|t_{\nu(i)}-t_{\nu(j)}|$
between different times sufficiently large, we can always force upon
the supports of Weyl operators at equal times a condition as
in~(4.20). Otherwise, the expectation would vanish. But then $\m=\o$, 
which is impossible.
\QED

The result of above shows that condition~(3.8) may hold in systems which are
weakly, but not strongly clustering. Therefore, quantized toral automorphisms
$(\A,\Theta,\phi)$ with deformation  parameter $\theta\notin\Z$ (see
Corollary~4) fall in the class of dynamical systems whose time-asymptotic
fluctuations can be handled via Proposition~3, by means of pair partitions
only. Condition~(3.12) obviously does not hold for any $\beta_r$ since its
equivalent version~(4.22) is easily violated. Taking into account that, when
$\beta_r\neq 0$, the asymptotic states  $\phi_\infty$ are not permutation
invariant,  we can summarize our findings for the previous class of  quantized
automorphisms of the torus in

\noindent
{\bf Corollary~6.}
{\it 
 When the deformation parameter $\theta$ is such that
 $\lim_t\beta(t)\theta/(\lambda^2-1)\ \mod 1$ equals
 $\beta_r$ with $\beta_r$ as in
 Proposition~5, the statistics of time-asymptotic
 fluctuations of quantized hyperbolic automorphisms 
 of the torus $(\A,\Theta,\phi)$ is as follows.
 For $\beta_r=0$, the fluctuations~(3.7) of centred observables $W_\theta(f)$
 are Gaussian random variables with variance $\sigma=\|f\|_2^2$.
 For $\beta_r\ne 0$, they obey a distribution law with vanishing odd moments and
 even moments $M_{2n}$ given by
 $$ 
   M_{2n}:=\lim_N \phi_\infty\biggl(F_N\Bigl(W_\theta(f)^{2n}\Bigr)\biggr)=
   {1\over n!} {\sum_\nu}^{(2)} \phi_\infty\Bigl(W_\theta(f)_{\nu(1)} 
W_\theta(f)_{\nu(2)}
   \cdots W_\theta(f)_{\nu(2n)}\Bigr)\,    
 $$
 where the sum is over all pair partitions of $\{1,2, \ldots, n\}$.
}

\noindent
{\bf Proof:}
 The statement follows from Propositions~3 and~11 and from Corollary~5.
\QED
\vfill\eject

\noindent
{\bf Acknowledgements:} 

F.B. acknowledges financial support from the Onderzoeksfonds K.U.Leuven
F/97/60 and the Italian I.N.F.N. and M. De Cock acknowledges financial
support from FWO-project G.0239.96.

{\parindent=0pt
\noindent
{\bf References}
\medskip

[ABDF]
Andries~J., Benatti~F., De~Cock~M. and Fannes~M.:
Multi-time correlations in relaxing quantum dynamical systems.
Preprint KUL-TF-98/48

[AGL]
Aubry~S., Godr\`eche~C., Luck~J.M.:
Scaling properties of a structure intermediate between quasiperiodic and
random.
J.\ Stat.\ Phys.\ {\bf 51}, 1033--1075 (1988)

[BV]
Balazs~N.L., Voros~A.:
The quantized baker's transformation.
Ann.\ Phys.\ {\bf 190}, 1--31 (1989)

[BF]
Benatti~F., Fannes~M.:
{\it Statistics of quantum chaos},
J.\ Phys.\ A  (in press)

[BNS]
Benatti~F., Narnhofer~H., Sewell~G.L.:
A non-commutative version of the Arnold cat map.
Lett.\ Math.\ Phys.\ {\bf 21}, 157--192 (1991)

[BBTV]
Berry~M.V., Balazs~N.L., Tabor~M., Voros~E.:
Quantum maps.
Ann.\ Phys.\ {\bf 122}, 26--63 (1979) 

[BKS]
Bozejko~M., K\"ummerer~B., Speicher~R.:
$q$-Gaussian processes: Non-commutative and classical aspects.
Commun.\ Math.\ Phys.\ {\bf 185}, 129--154 (1997)

[BS]
Bozejko~M., Speicher~R.:
An example of generalized Brownian motion.
Commun.\ Math.\ Phys.\ {\bf 137}, 519--531 (1991)  

[CC]
Casati~G., Chirikov~B.V.: 
{\it Quantum Chaos}. 
Cambridge: Cambridge University Press, 1995

[D]
Degli Esposti~M.:
Quantization of the orientation preserving automorphisms of the torus.
Ann.\ Inst.\ Henri Poincar\'e {\bf 58}, 323--341 (1993)

[HKS]
Haake~F., Kus~M., Scharf~R.:
Classical and quantum chaos for a kicked top.
Z.\ Phys.\ B{\bf 65}, 381--395 (1987)

[N]
Narnhofer~H.: Quantized Arnold cat maps can be entropic K-systems. J. Math.
Phys. {\bf 33}, 1502-1510 (1992)

[NT1]
Narnhofer~H., Thirring~W.:
Mixing properties of quantum systems.
J.\ Stat.\ Phys.\ {\bf 57}, 811--825 (1989)

[NT2]
Narnhofer~H., Thirring~W.:
C$^*$-Dynamical Systems that are Asymptotically Highly Anticommutative.
Lett.\ Math.\ Phys.\ {\bf 35}, 145--154 (1995)

[S]
Speicher~R.:
Generalized statistics of macroscopic fields.
Lett.\ Math.\ Phys.\ {\bf 27}, 97--104 (1993) 

[SvW]
Speicher~R., von~Waldenfels~W.:
{\it A general central limit theorem and invariance principle},
Quantum Probability and Related Topics IX, 371--387 
Singapore: World Scientific, 1994

[vLM]
van~Leeuwen~H., Maassen~H.:
A $q$-deformation of the Gauss distribution.
J.\ Math.\ Phys.\ {\bf 36}, 4743--4756 (1996) 

[VDN]
Voiculescu~D.V., Dykema~K.J., Nica~A.: 
{\it Free Random Variables}. 
Providence, RI: AMS 1992  
}
\bye